\documentclass[a4paper,conference]{IEEEtran}

\everymath{\displaystyle}

\ifCLASSINFOpdf
\else
\fi
\usepackage{amssymb, amsfonts, amsbsy, amsthm, latexsym, color}
\usepackage{url}
\usepackage[cmex10]{amsmath}
\everymath{\displaystyle}

\newtheorem{theorem}{Theorem}[section]

\newtheorem{lemma}[theorem]{Lemma}
\newtheorem{proposition}[theorem]{Proposition}

\newtheorem{definition}{Definition}

\newtheorem{remark}{Remark}[section]

\newcommand{\remove}[1]{}

\newcommand{\C}{{\mathbb{C}}}

\newcommand{\CHI}{\hbox{\raise .4ex \hbox{$\chi$}}}

\newcommand{\nsp}{$\text{NSP}$}

\newcommand{\nspD}{$D$-\nsp}
\newcommand{\snspD}{$D$-S\nsp}

\usepackage{amsmath, amssymb, amsfonts, amsbsy, amsthm, latexsym}

\begin{document}
%
\title{A null space property approach to compressed sensing with frames}


%

\author{\IEEEauthorblockN{\large Xuemei Chen}
\IEEEauthorblockA{Department of Mathematics\\
University of Maryland, College Park\\
Email: xuemeic@math.umd.edu}
\and

\IEEEauthorblockN{\large Haichao Wang}
\IEEEauthorblockA{Department of Mathematics\\
U C Davis\\
Email: hchwang@ucdavis.edu}

\and

\IEEEauthorblockN{\large Rongrong Wang}
\IEEEauthorblockA{Department of Mathematics\\
University of Maryland, College Park\\
Email:  rongwang@math.umd.edu}

}



%


\maketitle

\begin{abstract}
An interesting topic in compressive sensing concerns problems of sensing and recovering signals with sparse representations in a dictionary. In this note, we study conditions of sensing matrices $A$ for the $\ell^1$-synthesis method to accurately recover sparse, or nearly sparse signals in a given dictionary $D$. In particular, we propose a dictionary based null space property (\nspD) which, to the best of our knowledge, is the first sufficient and necessary condition for the success of the $\ell^1$ recovery. This new property is then utilized to detect some of those dictionaries whose sparse families cannot be compressed universally. Moreover, when the dictionary is of full spark, we show that $AD$ being \nsp, which is  well-known to be only sufficient for stable recovery via $\ell^1$-synthesis method, is necessary as well.

\end{abstract} 


%
\IEEEpeerreviewmaketitle

\section{Introduction}
Compressed sensing concerns the problem of recovering a sparse signal $x_0\in \C^d$ from its undersampled linear measurements $y=A x_0\in\C^m$, where the number of measurements $m$ is usually much less than the ambient dimension $d$. A vector is said to be \emph{$k$-sparse} if it has at most $k$ nonzero entries. The following  linear optimization algorithm, also known as the Basis Pursuit, can reconstruct $x_0$ efficiently from a perturbed observation $y=Ax_0+w$ where $\|w\|_2\leq\epsilon$~\cite{LqSim}\cite{Stab}:
\begin{equation}\label{equ_Pe}
\hat{x}=\arg\min_{x\in R^d} \|x\|_1, \ \ \text{subject to  } \|y-A x\|_2\leq \epsilon .
\end{equation}

A primary task of compressed sensing is to choose appropriate sensing matrix $A$ in order to achieve good performance of \eqref{equ_Pe}.  A matrix $A$ is said to have the \emph{Restricted isometry property (RIP)} with order $k$ if 
\begin{equation}\label{eq:RIP}
    (1-\delta)\|x\|_{2}^2\le\|A x\|_{2}^2\le
               (1+\delta)\|x\|_{2}^2
  \end{equation}
  for any $k$-sparse vectors $x$.
 RIP is shown to provide stable reconstruction of approximately sparse signals via \eqref{equ_Pe}~\cite{Decode}\cite{LqSim}. Moreover, many random matrices satisfy RIP with high probability \cite{Near, RV08}.
  A matrix $A$ is said to have the \emph{Null space property of order $k$ ($k$-NSP)} if 
\[
\forall v\in \ker A \backslash \{0\}, \ \forall |T|\leq k,\ \ \ \|v_T\|_1<\|v_{T^c}\|_1.
\] NSP is known as a characterization of uniqueness of problem \eqref{equ_Pe} when there is no noise~\cite{Gribonval}. It has also been proven that the NSP matrices admit a similar stability result as RIP does except that the constants may be larger~\cite{ACP11}.

A recent direction of interest in compressed sensing concerns problems where signals 
are sparse in an overcomplete dictionary $D$ instead of a basis,
see \cite{Dic_Candes, DMP1, Gribonval, LML12, ACP11, CoDic, Elad13}.  This is motivated by the widespread use of overcomplete dictionaries 
in signal processing and data analysis. Many signals naturally possess sparse frame coefficients, such as images consisting of curves (curvelet frame). In addition, the greater flexibility and stability of frames make them preferable for practical purposes in order to compensate the imperfectness of the measurements.
In this setting, the signal $x_0 \in \C^d$ can be represented as $x_0=Dz_0,$ where 
$z_0 $ is $k$-sparse and $D$ is a $d \times n$ matrix with $n\geq d$.  The columns of $D$ may be thought of as an
overcomplete frame or dictionary for $\C^d$.
The linear measurements are $y=Ax_0$.

A natual way to recover $x_0$ from $y$ is first solving 
\begin{equation}\label{equ_PD}
\hat{z}=\arg\min_{z\in R^n} \|z\|_1, \ \ \text{subject to  } y=AD z .
\end{equation}
for the sparse coefficients $\hat z$, then synthesizing it to obtain $\hat x$, i.e., $\hat x=D\hat z$. The resulting method is therefore called $\ell^1$-synthesis or synthesis based method~\cite{LML12, DMP1}. Since we are only seeking the recovery of $x_0$, we say the $\ell^1$-synthesis method \eqref{equ_PD} is \emph{successful} when every minimizer $\hat z$ of \eqref{equ_PD} satisfies $D\hat z=x_0$. 

 In the case when the measurements are perturbed, we naturally solve the following:  
\begin{equation}\label{equ_PDe}
\hat{z}=\arg\min_{z\in R^n} \|z\|_1, \ \ \text{subject to  } \|y-AD z\|\leq \epsilon .
\end{equation}

The work in \cite{DMP1} established conditions on $A$ and $D$ to make the compound $AD$ satisfy RIP. However, as pointed in \cite{Dic_Candes, LML12}, forcing $AD$ to satisfy RIP or even the weaker NSP implies the exact recovery of both $z_0$ and $x_0$, which is unnecessary if we only care about obtaining a good estimate of $x_0$. In particular, if $D$ is perfectly correlated (has two identical columns), then there are infinitely many minimizers of \eqref{equ_PD} that may be assigned to $\hat{z}$, but all of them lead to the true signal $x_0$. 
It seems reasonable to expect that similar result may hold in the case of highly correlated dictionaries, since they are only a small perturbation away from the perfectly correlated ones.

\subsection{Overview and main results}
In this paper, we generalize the ordinary null space property to the dictionary case (\nspD), and prove in Theorem \ref{thm_iff} that this new condition is equivalent to the accurate recovery of sparse signals in dictionaries via $\ell^1$-synthesis. 
Moreover, a stability result is given in Theorem \ref{thm_sta}. 
To the best of our knowledge, these results are the first characterization of compressed sensing with dictionaries via $\ell^1$-synthesis approach.

Section \ref{sec_adm} studies more properties of \nspD, and shows that $A$ has \nspD\ is equivalent to $AD$ has NSP as long as $D$ is of full spark (every $d$ columns of $D$ are linearly independent).
As a consequence, under the full spark assumption, 
the $\ell^1$-synthesis method cannot accurately recover the signals without accurate recoveries of their sparse representations,  therefore an incoherent dictionary is needed under this circumstance.

All proofs of the theorems presented can be found in \cite{CWW13}, while some proofs are provided here.
%
%
%
%

\section{A sufficient and necessary condition for noiseless sparse recovery}
In this section, we develop a sufficient and necessary condition for the success of $\ell^1$-synthesis method \eqref{equ_PD}.
We show that the following property on $A$ is a necessary and sufficient condition for successfully recovering all signals in $D\Sigma_s$ via \eqref{equ_PD}, where $D\Sigma_k=\{x: \exists\ z, \text{ such that } x=Dz, \|z\|_0\leq k\}$ is the set of signals that have $k$-sparse representations in $D$.
\begin{definition}[Null space property of a dictionary $D$ ($D$-NSP)]
Fix a dictionary $D\in \C^{d,n}$, a matrix $A\in \C^{m,d}$ is said to satisfy the $D$-NSP of order $k$ ($k$-\nspD) if for any index set $T$ with $|T|\leq k$, and any $v\in D^{-1}(\ker A\backslash\{0\})$, there exists $u\in\ker D$, such that 
\begin{equation}\label{eq: DNSP}\|v_T+u\|_1<\|v_{T^c}\|_1.\end{equation}
\end{definition}
\begin{theorem}\label{thm_iff}
\nspD\ is a necessary and sufficient condition for $\ell^1$-synthesis \eqref{equ_PD} to successfully recover all signals in the set $D\Sigma_k$. 
\end{theorem}
\proof Necessary part. We need to show that, if from measurements taken by a sensing matrix $A$, $\ell^1$-synthesis is successful in recovering all signals in $D\Sigma_k$, then $A$ must be $k$-\nspD. 

For any $v\in D^{-1}(\ker A/\{0\})$ and any index set $T$ with $|T|=k$, we define $x_0=Dv_T$ be a signal in $D\Sigma_k$, $y=Ax_0$ be its measurements, and let $\hat{x}$, $\hat{z}$ be the reconstructed signal and its coefficients from $y$ via \eqref{equ_PD}. If $\ell^1$-synthesis is successful for all signals in $D\Sigma_k$, then we must have $\hat{x}=x_0$, and so $\hat{z}=v_T+u$ with some $u\in \ker D$. 

Observe that $v_T-v$ is also feasible to \eqref{equ_PD}, but it is not a minimizer since it cannot be representated in the form of
$v_T+u$ with any $u\in \ker D$. Therefore, its $\ell_1$ norm is strictly greater than that of $\hat{z}$:
$$\|v_T+u\|_1<\|v_T-v\|_1=\|v_{T^c}\|_1,$$ implying $A$ is $k$-\nspD.

Sufficient part. Assuming $A$ is $k$-\nspD, we will show that the $\ell_1$ synthesis can recover all signals $ x \in D\Sigma_k$ from $y=Ax$. Suppose to the contrary that there exists an $x_0=Dz_0\in D\Sigma_k$, such that its reconstruction $\hat{x}=D\hat{z}$ is wrong. Then we must have $v:=z_0-\hat z\in D^{-1}(\ker A/\{0\})$. Let $T$ be the support of $z_0$, by \nspD, therefore there exists a $u\in \ker D$, such that $\|v_T+u\|_1<\|v_{T^c}\|_1$, i.e., $\|z_0-\hat z_T+u\|_1<\|\hat z_{T^c}\|_1$. Hence,
$$\|z_0+u\|_1\leq\|z_0-\hat z_T+u\|_1+\|\hat z_T\|_1<\|\hat z_{T^c}\|_1+\|\hat z_T\|_1=\|\hat z\|_1.$$ 
This is a contradicts to the assumption that $\hat z$ is a minimizer.\qed

Notice when $D$ is the canonical basis of $\C^d$, the \nspD\ is reduced to the normal \nsp\ with the same order. In other words, \nspD\ is a generalization of \nsp\ for the dictionary case. It is, however, a nontrivial generalization. 

The intuition of \nspD\ rises from the fact that we are only interested in recovering $x_0$ instead of the representation $z_0$. As long as the minimizer $\hat z$ lies in the affine plane $z_0+\ker D$, our reconstruction is a success.

\section{D-NSP based stability analysis}\label{sec_sta}
It is known that the \nsp\ is a sufficient and necessary condition not only for the sparse and noiseless recovery, but also for compressible signals with noisy measurement~\cite{ACP11, Sun11}. However, the stability analysis of \nsp~\cite{ACP11} cannot be easily generalized to our case because essentially
we need the function $f(v)=(\|v_{T^c}\|_1-\|v_T+u\|_1)/\|Dv\|_2$ to be bounded away from zero. In the basis case, we have knowledge of $f(v)$ on a compact set, and consequently the extreme value theorem can be applied to prove the exisitence of a positive lower bound. In our case we do not have a compact set, therefore other constructions to overcome this difficulty is necessary. 

\begin{definition}[Strong null space property of a dictionary $D$ (\snspD)]
A sensing matrix $A$ is said to have the strong null space property with respect to $D$ of order $k$ ($k$-$D$-SNSP) if there is a positive constant $c$ such that for any index set $T$ with $|T|\leq k$, and any $v\in \ker (AD)$, there exists $u\in\ker D$, such that 
\begin{equation}\label{equ_nsp}
\|v_{T^c}\|_1-\|v_T+u\|_1\geq c\|Dv\|_2
\end{equation}
\end{definition}

\snspD\ is a stronger assumption than \nspD\ by definition. We prove that under this assumption, the $\ell^1$-synthesis recovery is stable with respect to perturbations on the measurement vector $y$.

\begin{theorem}\label{thm_sta}
If $A$ is $k$-\snspD, then any solution $\hat z$ of problem \eqref{equ_PDe} satisfies
$$\|D\hat z-x_0\|_2\leq C_1\sigma_k(z_0)+C_2\epsilon.$$
where $\sigma_k(z_0)$ denotes the $\ell^1$ residue of the best $k$-term approximation to $z_0$, $C_1$, $C_2$ are constants dependent on $n$, the constant $c$ in \eqref{equ_nsp}, the minimum singular values of A and D, but not on $x_0$. 
\end{theorem}

\proof Let $x_0=Dz_0$ with $z_0$ being an $k$-sparse representation of $x_0$.
Let $h=D(\hat{z}-z_0)$, and decompose it as $h=Dw+\eta$ where $Dw\in\ker A$, $\eta\in \ker A^{\perp}$. It is easy to show that $\|\eta\|_2\leq \frac{1}{\nu_A}\|Ah\|_2\leq\frac{2\epsilon}{\nu_A}$ with $\nu_A$ being the smallest singular value of $A$.

Define $\xi=D^T(DD^T)^{-1}\eta$, then $\eta=D\xi$, and
\begin{equation}\label{equ_xi}
\|\xi\|_2\leq\frac{1}{\nu_D}\|\eta\|_2\leq\frac{2}{\nu_A\nu_D}\epsilon.
\end{equation}
Moreover, by our setting, $D(\hat{z}-z_0)=h=D(w+\xi)$, and therefore $\hat{z}-z_0=w+\xi+u_1$ with some $u_1\in \ker D$.

Let $v=w+u_1$, then $\hat{z}-z_0=v+\xi $ and $v\in \ker(AD)$. By the assumption of \snspD, there exists a $u\in\ker D$ such that \eqref{equ_nsp} holds for $u$ and $v$. Therefore,
\begin{align}\notag
&\|v+z_{0,T}\|_1-\|-u+z_{0,T}\|_1\\\notag
= &\|v_{T^c}\|_1+\|v_T+z_{0,T}\|_1-\|-u_T+z_{0,T}\|_1-\|u_{T^c}\|_1\\\notag
\geq &\|v_{T^c}\|_1-\|v_T+u_T\|-\|u_{T^c}\|_1\\\label{equ1}
=&\|v_{T^c}\|_1-\|v_T+u\|_1\geq c\|Dv\|_2
\end{align}
On the other hand, from the fact that $\hat{z}$ is a minimizer, we have
\begin{eqnarray*}
&&\|-u+z_{0,T}\|_1+\|z_{0,T^c}\|_1\geq\|-u+z_0\|_1=\|\hat{z}\|_1 \\ && \geq\|v+z_0+\xi\|_1 \geq\|v+z_0\|_1-\|\xi\|_1
\\ &&\geq\|v+z_{0,T}\|_1-\|z_{0,T^c}\|_1-\|\xi\|_1.
\end{eqnarray*}
Rearrange the above inequality, we will obtain
\begin{equation}\label{equ2}
\|v+z_{0,T}\|_1-\|-u+z_{0,T}\|_1\leq 2\|z_{0,T^c}\|_1+\|\xi\|_1.
\end{equation}
Combining \eqref{equ1} and \eqref{equ2}, we get 
\begin{equation}\label{equ_dv}
\|Dv\|_2\leq\frac{2}{c}\|z_{0,T^c}\|_1+\frac{1}{c}\|\xi\|_1\leq\frac{2}{c}\|z_{0,T^c}\|_1+\frac{\sqrt{n}}{c}\|\xi\|_2
\end{equation}
In the end, using \eqref{equ_dv} and \eqref{equ_xi},
\begin{align*}
\|h\|_2&=\|Dv+D\xi\|_2=\|Dv+\eta\|_2\leq\|Dv\|_2+\|\eta\|_2\\
&\leq\frac{2}{c}\|z_{0,T^c}\|_1+\frac{\sqrt{n}}{c}\|\xi\|_2+\frac{1}{\nu_A}2\epsilon\\
&\leq\frac{2}{c}\|z_{0,T^c}\|_1+\frac{2\sqrt{n}}{c\nu_A\nu_D}\epsilon+\frac{1}{\nu_A}2\epsilon.
\end{align*}
\qed

It is natural to ask how much stronger this new assumption is than \nspD. We address this question partially in the next section.



\section{A further study of \nspD\ and admissible dictionaries}\label{sec_adm}
This section explores the two assumptions \nspD\ and \snspD\ further for the purpose of answering the following important questions:
 What kind of dictionaries will allow sensing matrices $A$ with few measurements to satisfy \nspD? How to find those sensing matrices given a dictionary? 

We call a $d\times n$ dictionary $D$ \emph{$k$-admissible} if there exists a measurement matrix  $A\in \C^{m,d}$ with $m<d$ such that $A$ is $k$-\nspD. We call $D$ \emph{inadmissible} if $D$ is not $k$-admissible for any $k\geq 2$. Intuitively speaking, $D$ is not $k$-admissible means that $D\Sigma_k$ cannot be universally compressed by any linear matrix $A$.

The following proposition shows that adding repeated columns to the dictionary $D$ will not affect admissibility. This is quite intuitive since we do not change the set $D\Sigma_k$ during this procedure, and we only care about recovering the signal $x_0$ rather than the representation $z_0$.

\begin{proposition}\label{prop_repeat}
Let $D\in \C^{d,n}$, and let $I$ be any index set $I\subset \{1,...,n\}$. Define $\widetilde{D}=[D,D_I]$, then for any sensing matrix $A\in \C^{m,n}$, we have $A$ is \nspD\  if and only if $A$ is $\widetilde{D}$-\nsp.
\end{proposition}

Proposition \ref{prop_repeat} states that a perfectly correlated dictionary $D$ does not get in the way of the reconstruction of signals. It is only natural to ask whether this is still the case for a highly coherent dictionary. We answer this question partially by showing a class of highly correlated dictionaries is inadmissible. Moreover, equivalent conditions of \nspD\ is given in Section \ref{sec_nsp} under the assumption that $D$ is of full spark.
\subsection{A Class of inadmissible matrices}

The following theorem constructs a class of inadmissible matrices with a one dimensional kernel.
\begin{theorem}\label{thm_example}
Given 
 an orthonormal basis $\Phi=[\phi_1,...,\phi_d]$. Let $H=\bigcup\limits_{j=1}^d\text{span}\{\phi_{i}\}_{i=1,i\neq j}^d$ be the union of the hyperplanes spanned by every combination of $d-1$ columns of $\Phi$.  Then there exists a small constant $r_0$ such that for every $v\in B(\phi_1,r_0)\backslash H$ where $ B(\phi_1,r_0)$ is the ball centered at $\phi_1$ with radius $r_0$, 
  $D=[\Phi,v]\in \C^{d,d+1}$ is inadmissible.
\end{theorem}

We need the following lemma for the proof of this Theorem.

\begin{lemma}\label{contex}
Suppose $D$ is a $d\times (d+1)$ dictionary. If there exist $T\subset\{1,...,d+1\}$ with $|T|\geq 2$ such that any vector $u\in\ker D\backslash \{0\}$ satisfies
\begin{enumerate}
\item[1.]\label{item1} $\|u_T\|_1>\|u_{T^c}\|_1$, and
\item[2.]\label{item2} $T^c\subset \text{supp}(u)$,
\end{enumerate} 
Then $D$ cannot be $|T|$-admissible.
\end{lemma}
For any vector $w\in \mathbb{C}^n $, we define $\|w\|_{\min}=\min_{1\leq i\leq n}\{|w_i|\neq0\}$ to be the minimum magnitude in $w$.
\proof
Assume that the dictionary $D$ defined in Lemma \ref{contex} is $|T|$-admissible, we will show how this leads to a contradiction.  

Since $D$ is admissible, then there exists at least one $A$ that is $k$-\nspD. Pick one of them, and fix a $v_0\in D^{-1}(\text{ker}(A)\backslash\{0\})$. Define $\alpha= 2\|v_0\|_{\infty}/\|u\|_{\min}$. Now that $v_0+\alpha u, -v_0+\alpha u \in D^{-1}(\text{ker}(A)\backslash\{0\})$, by the definition of \nspD, there exist $c_1, c_2 \in \mathbb{C}$ such that
\begin{equation}\label{eq_1}
\|v_T+\alpha u_T-c_1 u\|_1<\|v_{T^c}+\alpha u_{T^c}\|_1,
\end{equation}
and
\begin{equation}\label{eq_2}
\|-v_T+\alpha u_T-c_2 u\|_1<\|-v_{T^c}+\alpha u_{T^c}\|_1.
\end{equation}
Therefore,
\begin{eqnarray}
&&2\alpha\|u_{T^c}\|_1 \label{eq:com}\\&=&\|v_{T^c}+\alpha u_{T^c}\| _1+\|-v_{T^c}+\alpha u_{T^c}\|_1 \label{eq:assump2}\\
&>&\|v_T+\alpha u_T-c_1 u\|_1+\|-v_T+\alpha u_T-c_2 u\|_1\label{eq:add}\\
&=&\|v_T+(\alpha-c_1) u_T\|_1+|c_1|\| u_{T^c}\|_1\notag\\&+&\|-v_T+(\alpha-c_2) u_T\|_1+|c_2|\| u_{T^c}\|_1 \notag\\
&\geq & |2\alpha-c_1-c_2|\|u_T\|_1+(|c_1|+|c_2|)\|u_{T^c}\|_1, \label{eq:last}
\end{eqnarray}
where \eqref{eq:assump2} follows from our assumption on $\alpha$ and Assumption 2, while \eqref{eq:add} from adding \eqref{eq_1} and \eqref{eq_2}. Combining \eqref{eq:com} and \eqref{eq:last} to get
\[
\|u_T\|_1 <\|u_{T^c}\|_1.
\]
This is a contradiction to Assumption 1 of Lemma \ref{contex}.
\qed

\vspace{0.1in}

\emph{Proof of Theorem \ref{thm_example}:} 
Notice that $\text{ker}(D)=\text{span}\{u\}$ with $u=(a^T,-1)$.
Let $T$ be an index set with $|T|\geq2$ such that $\{1, n+1\}\in T$. First, since $v\not\in H$, then $\langle v,\phi_i\rangle\neq 0$ for $i=1,...,d$ . This means that all coordinates of $u$ are nonzero, so Assumption 2 of Lemma \ref{contex} holds.
Second, we can pick $r_0$ small enough such that whenever $v\in B(\phi_1,r)$, it holds $\|u_T\|_1>\|u_{T^c}\|_1$, so Assumption 1 is satisfied.

Applying Lemma \ref{contex} completes the proof.
\qed


We have constructed an example of inadmissible dictionaries of special sizes: $d\times (d+1)$. The following proposition asserts that this dictionary can be used to generate inadmissible dictionaries of arbitrary dimension by adding appropriate columns to it.
\begin{proposition}\label{prop_sub}
If $D=[B, v]$ where $B$ is a full rank $d\times (n-1)$ matrix and $v=B\alpha$ with $\|\alpha\|_1\leq 1$, then
 $A$ has \nspD\   implies that $A$ has $B$-\nsp\ with the same order $k$.
\end{proposition}


\subsection{The relation between \nspD\ and \nsp}\label{sec_nsp}

It is obvious that $AD$ satisfies NSP implies $A$ satisfies D-NSP, which explains why imposing RIP or incoherence conditions on $AD$ could be too strong and unnecessary. To explore how much room there is between these two conditions can possibly answer the question whether we can allow highly coherent dictionaries or not, since $AD$ being NSP will inevitably leads to the incoherence of $D$. Surprisingly enough, we show that whenever $D$ is of full spark, these two conditions are equivalent.

A dictionary is of full spark means every $d$ columns of this matrix are linearly independent. 
\begin{theorem}\label{thm_main}
The following conditions are equivalent under the assumption that $D$ is of full spark,
\begin{itemize}
\item $A$ is $k$-\nspD;
\item $AD$ is $k$-\nsp;
\item $A$ is $k$-\snspD;
\item For any $v\in \ker AD$, there exists a $u$ such that 
\[
\|v_T+u\|_1<\|v_{T^c}\|_1.
\] 
\end{itemize}
\end{theorem}


\begin{remark}
We comment that full spark is not a strong assumption on matrices. In fact, full spark matrices is dense in the space of matrices~\cite{BCM12}, and a large class of full spark Harmonic frames is also constructed in~\cite{BCM12}. 
\end{remark}
\begin{remark}
Earlier we mentioned that we only care about recovering the signals $x$ and allow the recovery of their representations $z$ to be wrong. Theorem \ref{thm_main} tells us that when the dictionary is of full spark this requirement is actually not any looser than requiring both signals and their representations to be recovered. In spite of being negative, this result is quite important, since it has been largely thought that the opposite is true.


\end{remark}

Like the RIP, NSP is essentially an incoherence property of a matrix. Hence a highly coherent dictionary $D$ cannot be NSP, nor can the composite $AD$ be, because whichever vector in $\ker D$ that fails to satisfy NSP, is also contained in $\ker (AD)$. Consequently, the equivalence of the first two items in Theorem \ref{thm_main} implies that if a highly coherent $D$ is also full spark, then it must be inadmissible. 

 Perfectly coherent dictionaries are not full spark, so they can be and many of them are indeed admissible (Proposition \ref{prop_repeat}). However, if these dictionaries are perturbed a little bit, then no matter how small the perturbations are, with probability one, they will turn into highly coherent and full spark dictionaries and therefore become inadmissible. We conclude that admissibility is not stable with respect to perturbations.

\section*{Acknowledgment}
This research has been supported in part by Laboratory for Telecommunications Science (LTS) and by Defense Threat Reduction Agency HDTRA1-13-1-0015.



%
\bibliographystyle{plain}
\bibliography{citations_13_02_13}

\end{document}